\documentclass{article}

\usepackage{amsmath,graphicx,array}
\usepackage{dcolumn,soul}%
\usepackage{amsthm}
\usepackage[figuresright]{rotating}%
\usepackage{algorithm, algorithmicx, algpseudocode}
\usepackage{listings}%
\usepackage[sort]{natbib}
\bibpunct[, ]{(}{)}{;}{a}{}{,}
\newcommand{\GG}[1]{}
\RequirePackage{url}

\makeatletter
\def\uns{\ifmmode\,\else$\,$\fi}%

\makeatother
 
\title{Predicting Poverty}

\author{Paolo Verme\thanks{paolo.verme@unibo.it. The paper is published on the World Bank Economic Review, https://doi.org/10.1093/wber/lhae044. The author is grateful to Bo Pieter Johannes Andree, Olivier Dupriez, Aivin Vicquierra Solatorio, Lidia Ceriani, Hai-Anh Dang, David Newhouse, and two anonymous referees for comments on earlier versions.}}

\begin{document}
\thispagestyle{empty}
\maketitle

 \begin{abstract}
Poverty prediction models are used to address missing data issues in a variety of contexts such as poverty profiling, targeting with proxy-means tests, cross-survey imputations such as poverty mapping, top and bottom incomes studies, or vulnerability analyses. Based on the models used by this literature, this paper conducts a study by artificially corrupting data clear of missing incomes with different patterns and shares of missing incomes. It then compares the capacity of classic econometric and machine learning models to predict poverty under different scenarios with full information on observed and unobserved incomes, and the true counterfactual poverty rate. Random forest provides more consistent and accurate predictions under most but not all scenarios.
 \end{abstract}



\newpage

\section{Introduction}

The poverty rate, defined as the share of poor people in a given population, is an important indicator of well-being. ``End poverty in all its forms everywhere'' is the first of the the UN Sustainable Development Goals (SDGs) and the main instrument to monitor this goal is the poverty rate. It is used by International Financial Institutions (IFIs) for the global count of the poor, to classify countries according to their level of well-being, and allocate global financial resources. Estimates of poverty at the household level are also used by national and local governments to target populations in need of assistance and are a core instrument of social protection policies. An accurate estimate of poverty at the population or household level is a precondition for effective global, national and local welfare policies. 

Accurate poverty measurement is not a simple exercise. It is based on sample surveys that collect information on monetary indicators such as income, consumption or expenditure. It is therefore a sample based \textit{estimate} of the population poverty rate. These estimates suffer from a variety of errors including sampling errors, measurement errors, misreporting on the part of respondents or interviewers, and unit or item non-response. Once the data are collected, statistical agencies may also apply alterations to the data that can potentially impair proper statistical estimates such as top coding. No survey is exempted from at least some of these issues.

Among the various measurement issues described, unit and item non-response are particularly problematic in income surveys because of the share and type of missing incomes. The March supplement of the Current Population Survey (CPS) in the United States - the main instrument to measure poverty and inequality - suffered from a unit non-response rate that ranged between 4.3\% in 1979 to 16.7\% in 2018 with a quasi-linear increase over time (\citealp{Hlasny_2021}). If income alone is considered, the rate of item non-response for the CPS can be as large as 50\% depending on the income item considered (\citealp{MooreAl1997}). In 2019, household non-response rates among the 32 European countries that participated to the EU-SILC project - the main instrument to measure poverty and inequality in Europe - varied between 3.5\% in Turkey to 47.7\% in Ireland. These unit non-response rates have been increasing over time and do not take into account item (income) non-response.\footnote{See EU-SILC comparative reports (Annex 5) available at: https://circabc.europa.eu/faces/jsp/extension
/wai/navigation/container.jsp. No information specific to item non-response is available in these reports.} 

This problem does not apply to the US and Europe only but to any country engaged in the estimation of the poverty rate with incomes. The main global data repositories used for the measurement of poverty show that, worldwide, income is used to measure poverty more often than consumption. The Luxembourg Income Study provides net income for 50 countries and consumption for 25; PovCalNet, the World Bank data repository for the measurement of poverty, provides net income for 73 countries and consumption for 114; and the UNU-WIDER database provides net income for 163 countries and consumption for 66. Overall, high and middle-income countries tend to use income to measure poverty while low-income countries tend to use consumption.\footnote{See https://essd.copernicus.org/preprints/essd-2023-137/.}

Moreover, it has been shown repeatedly that missing incomes in surveys are typically Missing Not At Random (MNAR) and that they are concentrated on the tails of the income distribution (\citealp{AtkinsonAl2011}; \citealp{PikettySaez2003}; \citealp{Lillard_1986}, \citealp{Dalessio_2015}, \citealp{Bollinger_2019},  \citealp{Rubin_2020}; \citealp{Hlasny_2021}). This implies that any statistics estimated on observed incomes only is biased and that this bias should be expected to be particularly problematic for measures that are more sensitive to extreme values such as poverty and inequality. In countries that use incomes to measure the poverty rate, this rate can be severely biased because of the share and distribution of missing incomes. 

The problem with missing incomes cuts across several fields of the poverty measurement literature including poverty profiling, top and bottom incomes studies, targeting exercises, poverty mapping, vulnerability analyses, and various new methods based on big data and machine learning. The typical case of missing incomes is when some incomes are missing from the survey of interest. This has been of particular concern for practitioners working on poverty profiles in statistical agencies or international organizations, and for those scholars working on top and bottom incomes (\citealp{Cowell_1996b}, \citealp{AtkinsonAl2011}, \citealp{Jenkins2017}, \citealp{Hlasny_2021}), or vulnerability analyses (\citealp{Morduch1994}, \citealp{CalvoDercon2013}, \citealp{Verme_2016}). But there are many other cases where poverty is estimated out-of-survey with data that contain predictors of incomes but not incomes. This is the case of targeting exercises with proxy-means tests (\citealp{CoadyAl2004}, \citealp{BrownAl2018}, \citealp{Glewwe1991}, \citealp{BakerGrosh1994}) and cross-survey imputation exercises such as poverty mapping (\citealp{ElbersAl2003}, \citealp{TarozziDeaton2008}). More recently, scholars have started to use machine learning methods to estimate various measures of well-being with alternative data and predictors (\citealp{BlumenstockAl2015}, \citealp{AbelsonAl2014}, \citealp{NealAL2016}, \citealp{McBrideNichols2018}, \citealp{Bo2021}, \citealp{Aiken2022}, \citealp{Aiken2023}). Broadly speaking, missing incomes are an issue that affects any global or country study on poverty or incomes including seminal studies such as \cite{ChenRavallion2010}, \cite{Sala2006}, \cite{PikettySaez2003}, \cite{Burkhauser2023}, \cite{Deaton2005}, and many others.

Statistical biases due to income non-response are well known among statisticians and economists. Poverty specialists, statistical agencies, and economist more in general tend to experiment with alternative prediction methods to fill in for missing incomes such as single and multiple imputations, matching methods, replacing methods using moments of the income distribution, replacing methods based on various parametric functions, re-weighting methods based on the estimated probability of non-response, and others. These methods have been used extensively by the literature quoted above. However, in our knowledge, the relative performance of the poverty prediction models used by this literature has not been tested whereas it remains challenging to test the accuracy of each model with empirical data because the true poverty rate of the full distribution of incomes is not known due to the share and distribution of missing incomes. 

Building on the literature cited above, this paper compares the performance of classic econometric\footnote{By ``classic econometric'' we mean standard OLS and Maximum Likelihood models such as logit or probit models.} and machine learning models in predicting poverty with different missing observations shares and patterns and against the true poverty rate. This is done by generating a sample with no missing observations' bias and corrupting this sample with various shares and patterns of missing observations that mimic real world data. The performance of poverty prediction models can then be assessed with complete information on the full distribution of incomes, the true poverty rate, and the specific missing data patterns. The paper also provides a framework to compare classic econometric and machine learning models. 

The objective of the paper is to show how different classic econometric and machine learning models behave for predicting poverty when data, objective function (loss function), poverty lines, or various parameters and prediction strategies change. We are not striving to find the ultimate prediction model for the data at hand (which is a dummy data set) or causal relations, but understand how different prediction models respond to changes in these features using a dummy data set derived from real world data. The analysis is based on income data and the prediction models considered require income predictors to be observed. Results of this paper are not necessarily valid for other money metrics of well-being such as consumption or expenditure. 

Results show that the quality of poverty predictions and the choice of the optimal prediction model can dependent on the distribution of observed and unobserved incomes, the poverty line, the choice of objective function and policy preferences, the choice of models' parameters, and the use or non use of various optimization strategies. Random forest models are more robust than other models to variations in these features. This is due to the better ability of this model to predict incomes in the tails of income distributions, even with basic specifications. Other machine learning models can reach similar results but only with a full grid search and a substantial cost in terms of computational time.

The paper is organized as follows. The next section describes common missing data problems. Section 3 outlines how welfare economists have addressed this problem. Section 4 provides a consistent framework that can be used to compare classic econometric and machine learning models. Section 5 describes the data. Section 6 conducts a study with dummy data to compare the capacity of these models to predict the poverty rate accurately. Section 7 provides a series of robustness tests by varying data, parameters and preferences. Section 8 concludes by summarizing the main findings and providing some initial indications on how these models can be used effectively.

 \section{The distribution of missing data}
Addressing a missing data issue requires an understanding of the nature of missing data. Statisticians (\citealp{Rubin_1976}, \citealp{Rubin_2020}) distinguish between data Missing Completely At Random (MCAR), when there is no apparent law that regulates missing data; Missing At Random (MAR) where missing data of the outcome variable of interest are correlated with covariates of this outcome but not with the outcome itself (for example, when only men are not responding to income questions because of their gender, not their income), and Missing Not At Random (MNAR) where missing data of the outcome variable of interest are correlated with the outcome variable itself (for example, higher income households are less likely to respond to income questions in surveys because they have higher incomes). 

More formally and following \cite{Rubin_2020}, let $D_{ij}$ be the complete data matrix and $M_{ij}$ the indicator matrix representing missing observations where $i$ represent observations and $j$ represent variables. Then, the distribution of $m_i$ conditional on $d_{i}$ is $f_{M|D}(m_{i}|d_{i}, \phi)$ with $\phi$ being the unknown parameters of the function that relates $m$ to $d$. If $M_{ij}$ does not depend on $D_{ij}$, it is said that missing data are Missing Completely at Random (MCAR). Let now $d(0)_{i}$ be the components of $d_{i}$ that are observed for unit $i$, and $d(1)_{i}$ the components of $d_{i}$ that are missing for unit $i$. A less restrictive assumption than MCAR is that $m_{i}$ depends on $d_{i}$ only through the observed components $d(0)_{i}$. This case is defined as observations Missing At Random (MAR). Finally, the missing data pattern is called Missing Not At Random (MNAR) if the distribution of $m_{i}$ depends on $d_{i}$, which is only partially observed.

In the case of poverty measurement, the problem of non-randomness is particularly acute when poverty is measured with incomes, a standard practice in high and many middle-income countries. Missing incomes in surveys are known to be correlated with income itself in rich and poor countries alike (\citealp{AtkinsonAl2011}, \citealp{Hlasny_2018b}), and there is evidence that income non-responses are either an increasing function of income or are U-shaped with both lower and upper income households less likely to respond to questions in surveys ({\citealp{Lillard_1986}; \citealp{Bollinger_2019}; \citealp{Dalessio_2002}; \citealp{Dalessio_2015}). This fact makes income data MNAR and possibly also MAR, since some predictors of incomes are also likely to be associated with missing data. From a statistical standpoint, this is the most complex scenario for proper statistical estimates because the function that relates $m_{i}$ to $y_{i}$ (the $\phi$ parameters) is unknown. It is a scenario where the basic assumptions needed to use popular imputation methods such as multiple imputations are not met. 

In our knowledge, the only strand of the poverty literature that has focused on this problem is a string of papers that uses a GMM method to estimate the probability of non-response in sample. This method estimates the function that relates $m_{i}$ to $d_{i}$ (the $\phi$ parameters) addressing the fundamental missing data problem. Authors who used this method found income non-responses to be strongly associated with income (\citealp{Korinek_2006}; \citealp{Korinek_2007}; \citealp{Hlasny_2018b}; \citealp{Hlasny_2018}; \citealp{Hlasny_2021}). When one has a large share of missing incomes and no means to test their pattern, one should assume that missing incomes are MNAR and that estimating the poverty rate on observed values only is likely to bias this estimation significantly. Since the data necessary to estimate the probability of non-response are generally not available to researchers, it is essential to test how prediction models behave in the presence of non-response biases.

While the problem of missing incomes is generally discussed in the context of in-survey imputations, it is equally relevant for out-of-survey imputations as in targeting and cross-survey imputations for two reasons. One is that the original sample used for modeling incomes is also likely to suffer from missing incomes as most surveys do, which has implications for the predicted values out-of-survey. And second, even if all incomes are observed in the modeling sample, predicting out of sample amounts to predicting missing incomes for the totality of households (individuals) in the imputation sample. This can also be regarded as a missing data problem where 100\% of incomes are missing.

\section{How poverty economists address missing data issues to predict poverty}
There are several strands of the poverty measurement literature that treat the question of missing incomes with prediction models based on classic econometric methods: The literature on targeting and proxy-means testing, the literature on cross-survey imputations such as poverty mapping, the literature on top and bottom incomes, the literature on vulnerability to poverty, and the in-survey imputation literature. More recently, poverty economists have experimented with machine learning methods. We briefly review these strands of the  literature in this order.

The literature on proxy-means testing used for targeting makes extensive use of prediction methods to estimate poverty for households where income is non-observed (\citealp{CoadyAl2004}, \citealp{BrownAl2018}, \citealp{Glewwe1991}, \citealp{BakerGrosh1994}). The idea is that one can predict poverty using a restricted set of observed socio-economic characteristics avoiding in this way extensive and expensive surveys on income or consumption. In this case, the prediction model is built on an existing survey representative of the population of interest. A short survey is then administered to potential beneficiaries of welfare programs to collect data on key income predictors as identified by the model. This information is, in turn, used to predict poverty for individual households or assign a score that can rank households according to their level of well-being. This literature has used standard OLS or Logistic models for the prediction model. It implicitly assumes that missing data from the sample used for the prediction model are not problematic and that targeting beneficiaries are extracted randomly from the same population covered by the prediction model. A proper discussion of missing data in works on proxy-means testing is rarely seen but the models used are poverty prediction models covering households with missing incomes.

Cross-survey imputation methods have been developed to estimate poverty when income or consumption data are missing from the survey of interest but can potentially be estimated using other surveys representative of the same population and including incomes. One example is small areas estimations also referred to as ``poverty mapping''. The idea behind poverty mapping is to use poverty predictors extracted from censuses to predict poverty at the micro geographical level using the coefficients of a prediction model estimated with survey data that contain incomes at the macro geographical level (\citealp{ElbersAl2003}, \citealp{TarozziDeaton2008}). Although this literature has developed rather independently of the multiple imputation literature in statistics\footnote{\cite{ElbersAl2003} does not refer to the Rubin or Imbens literature while \cite{TarozziDeaton2008} refers to several of Rubin's papers but not to those that specifically addressed the cross-survey imputation question.}, it uses multiple imputation methods and builds on these methods to specifically address the question of reduced variance among predicted values in the context of continuous dependent variable models. Similar imputation methods for the purpose of predicting poverty have also been used across years (\citealp{DangAl2019}), different types of surveys such as consumption and labor force surveys (\citealp{DouidichAl2016}), or different types of data such as administrative and survey data (\citealp{DangVerme2019}). This literature has used continuous (\citealp{ElbersAl2003}) and categorical (\citealp{TarozziDeaton2008}) dependent variable models to predict poverty. As for the proxy-means tests literature, this literature rarely discusses missing data patterns but addresses the important question of the correct estimation of the variance of predicted values with OLS models.

The literature on top and bottom incomes has focused on the fact that incomes in the tails are under represented in surveys and that a correct estimation of inequality or poverty in any given country needs to address this issue (\citealp{AtkinsonAl2011}, \citealp{Jenkins2017}, \citealp{Hlasny_2021}, \citealp{Cowell1996}, \citealp{HlasnyAl2021}). This literature recognizes that missing observations are an increasing function of income and are, therefore, MNAR.\footnote{Interestingly, this literature rarely refers to MNAR data explicitly.} Several methods have been proposed to address this issue ranging from replacing incomes in the tails with observations extracted from theoretical distribution functions such as Pareto (\citealp{Cowell_1996b}, \citealp{Jenkins2017}), to replacing top incomes with data external to the survey such as tax data (\citealp{AtkinsonAl2011}), to reweighting observations using the inverse of the probability of non-response estimated from observed data (\citealp{Korinek_2007}, \citealp{Korinek_2006}, \citealp{Hlasny_2018b}). Replacing observations with theoretical distributions or external data can be effective when missing observations are almost exclusively on the tails of a distribution but these methods are less efficient when missing observations are located closer to central values. Reweighting methods are more indicated to estimate missing observations all along the distribution and they also have the distinct advantage of estimating the probability of non-response, which is the function that relates $m_{i}$ to $d_{i}$. However, in order to implement this method, one has to have non-response rates at a very disaggregated level, an information that is not always available to researchers. Unlike the proxy-means testing and cross-survey imputation literature, this literature focuses on the missing data question.  

Scholars working on vulnerability to poverty have also used prediction methods to gauge the probability of poverty in the future by simply estimating this probability with a OLS or Logit prediction model (\citealp{Morduch1994}; \citealp{CalvoDercon2013}; \citealp{Verme_2016}). This literature has not been particularly concerned with either missing items as the top incomes literature or the error term as for the cross-survey imputation literature. However, it is similar to the case where the cross-survey imputation methods are applied to surveys administered in different years, with the important difference that predictions are made in-survey and not out of-of-survey.  

All these strands of the literature may also use in-survey single or multiple imputations to estimate incomes for item non-response in the data used for modeling. This is where households are captured in the sample but do not reply to the income question. In this case, one can estimate incomes based on the other socio-economic characteristics observed with single or multiple imputation methods. This is also a standard practice used by practitioners working on poverty profiles.

More recently, machine learning methods have been used by economists to predict poverty with a variety of innovative data such as mobile phone (\citealp{BlumenstockAl2015}), satellite imagery and remote sensing data (\citealp{AbelsonAl2014}, \citealp{NealAL2016}), or for targeting the poor (\citealp{McBrideNichols2018}; \citealp{Aiken2023}). A global competition launched by the World Bank to predict poverty with machine learning algorithms provided some initial evidence on how these methods can help to improve on classic poverty prediction methods.\footnote{See details of this competition on https://www.drivendata.co/blog/poverty-winners/.} All these studies largely relied on standard ML methods including tree based methods, elastic nets, and neural networks, or deep learning methods. These are the ML methods considered by this paper.

\section{Baseline Framework for Comparing models}

\subsection{Three steps' predictions}

As shown in the previous section, one important distinction that the different strands of the poverty prediction literature share is the distinction between continuous and discrete (dichotomous) dependent variable models. These two types of models are applicable in the context of classic econometric and machine learning models lending themselves to be a useful framework to compare these different approaches to poverty predictions. This section clarifies the steps required to classify households into poor and non-poor households and the difference between these two sets of models.

To illustrate these differences, we use a simple OLS model based on a continuous income variable and a logit model based on a categorical binary variable that classifies the population into poor and non-poor statuses.\footnote{Note that one could use an OLS model with a binary dependent variable but this practice is rare.} In the remaining of the paper, we refer to the first model as the `income' (welfare) model and the second model as the `poverty' model, with both models leading to poverty predictions. Predicting household poverty with these two models requires three steps which we define as `Modeling', `Prediction' and `Classification' and are described as follows:

\textbf{Step 1 - Modeling}
\begin{equation}\label{eq:welf1}
W_{i}=\alpha+\beta_{1} X_{i}+\eta_{i}+\epsilon_{i} \\
\end{equation}
\begin{equation}\label{eq:pov1}
P_{i}=\delta+\gamma_{1} X_{i}+\nu_{i}+\psi_{i}
\end{equation}

where $i$ is the unit of observation (usually a household or an individual, household for short), $W_{i}$ = income, $P_{i}$ =poor where $P_{i}$ =1 if the unit is on or under the poverty line and $P_{i}=0$ otherwise, $X$ is a vector of household or individual characteristics, $\eta_{i}$ and $\nu_{i}$ are random errors and $\epsilon_{i}$ and $\psi_{i}$ are model fitting errors. 

The second step is the prediction of income or poverty based on the coefficients estimated under the modeling equations:

\textbf{Step 2 - Prediction}
\begin{equation}\label{eq:welf2}
\widehat{W_{i}}=\widehat{\beta_{1}} X_{i}+\tilde{\eta_{i}}+\tilde{\epsilon_{i}}
\end{equation}
\begin{equation}\label{eq:pov2}
\widehat{P_{i}}=\widehat{\gamma_{1}}X_{i}+\tilde{\nu_{i}}+\tilde{\psi_{i}}
\end{equation}

where $\widehat{W_{i}}$, $\widehat{P_{i}}$ are predicted income and poverty and $\tilde{\eta_{i}}, \tilde{\epsilon_{i}}, \tilde{\nu_{i}}, \tilde{\psi_{i}}$ are the estimated random and model fitting errors. Step 2 is the key step for addressing missing data issues. This is where missing incomes or poverty status are replaced with predicted values. 

The third and final step is to divide the population into estimated poor and non-poor groups. For this purpose, the welfare and poverty models critically differ in several important respects. Under the income model, the poverty line is used after the second step to separate the poor from the non-poor. Under the poverty model, the same poverty line is used to separate the poor from the non poor to construct the poor dichotomous variable in step 1 based on observed values. Once the probability of being poor is estimated for missing incomes in Step 2, a probability cut-point is used to separate the poor from the non poor. Therefore, Step 3 can be described as follows:

\textbf{Step 3 - Classification}
\begin{equation}\label{eq:welf3}
\begin{split}
if \; \widehat{W_{i}}\leq{z}: i=poor \\
else: \; i=nonpoor
\end{split}
\end{equation}
\begin{equation}\label{eq:pov3}
\begin{split}
if \;  \widehat{P_{i}}>prob^{*}: i=poor \\
else: \;  i=nonpoor
\end{split}
\end{equation}

where $z$ is the poverty line with ${W_{min}\leq z \leq W_{max}}$ and $prob^{*}$ is a probability cutpoint with $0\leq prob^{*} \leq 1$ that can be arbitrary or defined with some form of optimization criteria.

A second important difference between the welfare and poverty models is that the income model is typically estimated with an Ordinary Least Squares (OLS) estimator whereas the poverty model is estimated with a Logit or Probit maximum likelihood estimator. 

A third difference is that the income model produces income predictions whereas the poverty model produces probabilities of poverty predictions. One can easily turn the monetary predictions from the income model into probability of poverty predictions. In fact, for each poverty line $z={x_0,...,x_n}$ the probability of poverty of a household with income $x$ is 1-F(x). Therefore, we can express both models in terms of probabilities of being poor. However, in practice, scholars have used income or probability of poverty predictions depending on the model used. This implies that comparisons between the two models can only be made after the classification step.

Poverty predictions from both models can be improved after Step 2. The OLS income model produces a distribution of predicted values that is narrower than the true distribution. This is a statistical fact that has important implications for poverty predictions and that has induced scholars working on cross-survey imputations to propose specific solutions. Poverty predictions from the logit/probit models can also be improved by shifting the probability threshold in order to optimize the trade-off between different types of errors. This is done using Receiver-Operating Characteristic (ROC) curves and indexes initially introduced in clinical medicines but also used by poverty specialists (\citealp{Wodon_1997}; \citealp{Verme_2019}). These adjustments will be considered further in the paper.

The income and poverty models described above is what we refer to as `classic econometrics' models. We also consider three families of machine learning models: Decision Trees, Regularization, and Neural Networks. In particular, we use Random Forest, Elastic Nets, and Neural Networks with two hidden layers as representative choices of these families. As already discussed, these models are the most popular among economists. Regularization models rely on the same OLS and Logistic models described with the important difference of `regularization' as a method to shrink parameters. Random forest uses its own classification method based on entropy measures used to split the data in groups as homogeneous as possible and a random selection process for data and variables (bootstrap aggregation) to obtain optimal out-of-sample predictions. Neural networks can be seen as parametric functions such as OLS models with a very high number of parameters that are determined by the trial and error process in-built in the model. Details of these models are discussed in the calibration section further in the paper.

\subsection{Confusion matrix, errors, and objective functions}

All poverty prediction models illustrated above will result in true and false predictions that are best illustrated with a confusion matrix (also known as error matrix) resulting after Step 3 of the modeling exercise (Table \ref{table:Table 1 - ConfusionMatrix}). The matrix divides the population into four groups based on whether predictions are correct or not: True Positives (TP), True Negatives (TN), False Positives (FP) and False Negatives (FN). The primary objective of any classification exercise is to maximize TP and TN and minimize FP and FN. Incorrect classifications result in errors Type I and Type II. In the case of poverty predictions, Type I error refers to non-poor persons who are erroneously predicted as being poor. This error is also known as False Positive Rate (FPR), inclusion error, or leakage rate and is defined as FP/(FP+TN). Type II error refers to persons who are poor but are erroneously predicted to be non-poor. This error is also known as False Negative Rate (FNR), exclusion error, or undercoverage rate and is defined as FN/(FN+TP).  

\begin{center}
[Table \ref{table:Table 1 - ConfusionMatrix}]
\end{center}

To determine the objective function to optimize when predicting poverty, one has to be clear about what is observed and not observed, the type of error to minimize, and whether the objective is to optimize the estimation of poverty at the population or household level.

The first observation is that, with survey data, not all incomes are observed which implies that none of the values in the four cells of the confusion matrix is known. What is known after the final classification step is the number of predicted poor and predicted non-poor, or the sums of the two columns of the confusion matrix. In these cases and in order to test the quality of the prediction models, scholars split the available sample into a train and test samples, run the model on the train sample and then evaluate how the model performs on the test sample making the important assumption that the test sample is free of missing incomes' biases. Repeating this exercise several times with samples of different sizes provides a robustness test for the prediction model. In our case and as explained in the data section that follows, we start with a sample free of missing incomes biases which implies that all four cells of the confusion matrix are known.

The next question is the choice of objective function to optimize. Type I and Type II errors can be regarded as important from the perspective of an administrator of a poverty reduction program. Minimizing Type II error (exclusion error) is clearly more important from a poverty perspective. Excluding some of the poor has real consequences for well-being. But Type I error (inclusion error) may also be considered important if budgets are constrained, which is a common feature of poverty reduction programs worldwide. Including non-poor people is costly and diverts resources from the poor. How much importance should be given to each objective is, of course, a matter of \textit{preferences} and the trade-offs between the two objectives also depend on the relative \textit{cost} of inclusion or exclusion, which is case/country specific.
 
The third question and in the case of poverty predictions, the objective function to consider is different depending on whether one is interested in estimating the poverty rate as a population statistics (anonymous case), or estimating the poverty status correctly for each household (non-anonymous case). If the objective is to predict the poverty rate for the population, it is not essential to minimize both Type I and Type II errors. It is sufficient to minimize the difference between the true population poverty rate $P$ and the predicted poverty rate $\widetilde{P}$: 

\begin{equation}
min({P-\widetilde{P}})=min[(\widehat{P}+\epsilon)-\widehat{P}]=min(\epsilon)
\end{equation}

If we refer to the confusion matrix, this is equivalent to maximizing the sum of the true predictions ($max(TN+TP)$) or minimizing the sum of the false predictions ($min(FN+FP)$) irrespective of the actual TN or TP (FN or FP) values. In econometric terms, this amounts to minimizing the average \textit{model} error term for the population (not the \textit{idiosyncratic} error term which averages zero). In this case, and provided we are conducting an experiment where we know the true poverty rate, a possible test to evaluate the performance of the models is a means difference test between the true and predicted poverty rates. 

However, a means difference test between true and predicted poverty can only be conducted if the true poverty rate is known. In statistics and with survey data, prediction errors are usually evaluated with a range of indicators. Popular indicators for linear regression models are Mean Bias Error (MBE), Mean Square Error (MSE), Mean Absolute Error (MAE), or Root Mean Squared Error (RMSE), while a common approach to classification problems is to use some form of cross-entropy indicator. These indicators have been used to evaluate poverty prediction models but they cannot be used to compare continuous and discrete dependent variables models. They are also designed to evaluate the performance of regression models rather than the capacity to classify outcomes accurately and they are not measured against the true counterfactual (the true poverty rate in our case). 

Estimating the correct population poverty rate may not be sufficient if one has an interest in correctly estimating the poverty status of each household included in the sample. This complicates the objective function and the optimization process as we now have two elements to maximize (TN and TP) or minimize (FP and FN). We also need to attribute relative preferences to the two elements unless we consider the two elements equally valuable. One simple way to do that is to maximize the weighted sum of TP and TN as

\begin{equation}
max[a*TP+b*TN].
\end{equation}  

With $a$ and $b$ indicating preferences for TP and TN. In general, one would prefer to maximize TP and maximize coverage as opposed to maximizing TN and minimizing leakage. However, budget considerations may also be important and different policy makers may have different preferences for $a$ and $b$. 

In order to compare continuous and dichotomous dependent variable models and in order to clarify the weight given by the objective function to the four elements of the confusion matrix, one has to define the objective function in terms of the elements of the matrix. Following this strategy, different scholars have developed alternative objective functions such as the True Positive Rate, sensitivity or recall (TPR=TP/(FN+TP)), the True Negative Rate or specificity (TNR=TN/(TN+FP)), precision (TP/(TP+FP)) or the False Discovery Rate (FP/(TP+FP), and the accuracy ratio ($(TP+TN)/N$). The difference between these functions is simply the weight attributed to each of the four elements in the matrix. In a sense, they are different ways of expressing preferences for different types of errors. They also require knowledge of all values in the four cells of the confusion matrix. Throughout the paper, we will focus on $max(TP+TN)$ and $max[a*TP+b*TN]$ but we will also use these other functions to illustrate whether different preferences may lead to different choice of prediction model. 

\section{Data}
To observe the true poverty rate and measure the true prediction error, we generate a dummy data set free of missing data bias starting from real data. We first identified a publicly available real data set with an exceptionally low unit non-response rate, no item non-response, and information on non-response rates by geographical areas.\footnote{The survey we use is the 2007 round of the Enqu\^ete Nationale sur les Niveau de Vie des M\'enages (ENNVM) of Morocco which has a (unit) non-response rate of below 2\% given a full sample of 7,200 observations by design and a final surveyed sample of 7,062 observations. The survey has no item non-response and is also one of the few surveys for which it was possible to recover unit non-response rates by geographical areas (regions and urban/rural areas), which is essential to estimate the probability of non-response. Microdata and meta-data for the survey can be downloaded freely from the web (https://www.hcp.ma).} We then re-weighted observed incomes using the inverse of the probability of non-response correcting for any remaining bias due to unit non-response. This methodology is very well-known among statisticians (\citealp{Rosenbaum1987}; \citealp{Kim2007}) and widely used by economists working on top incomes (\citealp{Korinek_2007}; \citealp{HlasnyAl2021}). It has the distinctive advantage of being able to estimate the function that relates $m_{i}$ to $d_{i}$ (the $\phi$ parameters), estimate the probability of non-response, and returning a sample of incomes that is statistically equivalent in mass and representation to the original sample as intended by survey design. In other words, we start with real data which already have a very marginal problem with missing incomes, and exploit the availability of non-response rates by geographical area to re-weight observed incomes and obtain a sample which is expected to be statistically free from any income bias.

It is possible to think of other choices of data and methods to compare prediction models but none of these choices, in our view, would be as satisfactory. For example, one could generate an artificial income distribution using parametric functions that are well-known to represent incomes well such as the generalized beta distribution with two or more parameters. However, it would be almost impossible to generate artificially a set of predictors that could be used in the prediction model and that could approximate real data behavior. A second possibility is to take real data, superimpose a parametric distribution on incomes, and then extract missing incomes from this theoretical distribution. This is possible and a common approach among scholars working on top incomes, but it is unknown whether the chosen distribution would represent the data at hand well, and it is unknown the exact location of the missing incomes to extract. An alternative method is to use data external to the survey to complete the survey data such as tax or social security records, a method also popular among scholars working on top incomes. This has several problems. One is that the definitions of income are often different in the two sources of data. A second is that one should be able, as a minimum, to match the observed survey data with the respective external data before using external data to replace missing observations in the survey data. This is something that is very hard to do and something that we have not seen in published papers. Moreover, tax and social security data are also plagued by misreporting and under reporting, which carries the risk of introducing new biases.

An additional possibility is to conduct an experiment where non-responding households are tracked and encouraged to respond. If the survey administrators manage to recover all missing respondents, one could compare the original and final samples. A similar strategy has been followed in \cite{Blattman2014} although the authors opted to recover only a share of the missing respondents in their panel survey. This strategy has the potential to be an optimal strategy but recovering all non-respondents is very unlikely in practice, the two surveys cannot be administered at the same time, and incentives to reply to a second round of interviews may generate behavioral changes. A fifth possibility is simply to compare models using multiple data sets from different types of countries with different non-response rates. However, it is usually difficult to obtain from statistical agencies information on missing incomes, which makes it impossible to know the nature and distribution of missing incomes. Even where this information is available, with large shares of missing incomes, it is very hard to trust post-survey corrections with re-weighting or replacing methods. Finally and critically, with the exception of the case where survey administrators are able to recover all missing respondents, none of the methods described above dispose of a true counterfactual, the poverty rate that would have been estimated from the complete distribution of incomes with no missing values.  

As a measure of income, we use the log of income per capita calculated on households rather than individuals, and without sampling weights. This is done on purpose and for simplicity to avoid a discussion on adult equivalence scales and the use of survey weights in regressions, which is beyond the scope of this paper.\footnote{If the objective is not to estimate population statistics but study causation or prediction accuracy (as in our case) there are pros and cons in using sample weights and quite a bit of disagreement among econometricians on whether they should be used. Also, not all machine learning models allow for sampling weights or, if they do, they are a black box. It is unclear how they enter the algorithms, what effects they have on predictions, and whether different models treat them differently. As our objective was not to estimate population statistics for Morocco but compare the performance of prediction models, avoiding weights is a cleaner choice.} The set of independent variables is rather standard and includes age, age squared, gender, marital status, skills, occupation and working sector of the head of the household, and household size and location (urban/rural). All these variables are used in binary form except for age and household size and they are all fully observed. This set of independent variables is the same for all models in the paper. Summary statistics for all variables are in Table \ref{table:Table 2 - SummaryStatistics}.

\begin{center}
[Table \ref{table:Table 2 - SummaryStatistics}]
\end{center}

To compare models' performance when missing data patterns change, we corrupt our dummy data set mimicking eight missing data patterns: five MCAR selecting randomly different shares of missing data (5, 25, 50, 75, and 95\%), ``MAR pure'' meaning that we randomly selected 50\% of the sample conditional on one independent variable that is not correlated with income (we use working individuals in the secondary sector which we tested for independence of income), ``MAR-MNAR'' where we randomly selected 50\% of the sample conditional on a variable which is associated with income (we use household size$<$5), and ``MNARpure'' where we randomly selected 50\% of the sample conditional on income only (we use income$>$mean income). The most common and relevant case for this paper is MAR-MNAR whereas MNARpure is a rare case in empirical contexts. The resulting income distributions are plotted in Figure 1. This is, of course, a simple characterization of MAR and MNAR issues. Models that include a large set of regressors may exhibit more complex correlation patterns between missing incomes, observed incomes, and other independent variables.  

\begin{center}
[Figure 1]
\end{center}

Our approach has two additional advantages versus other possible methods that can be used to compare models. After corrupting the data, the resulting distributions are different in size and shape. Comparing models' performance across the eight data sets generated is, therefore, a test across different missing data patterns, but also a test across different shapes of the income distribution as if we were comparing different data sets. Since almost all income surveys contain some degree of missing incomes, we are mimicking - with real data - other types of income distributions found in surveys with the distinct advantage of knowing the type and distribution of missing values and the true counterfactual poverty rate. This is something that would not be possible by simply comparing the prediction models across different real data sets. Moreover, by removing different shares of observations randomly as we do with MCAR corruption patterns, we are explicitly creating train and test samples for in-sample and out-of-sample predictions of different sizes. Therefore, the models' prediction capacity is tested on out-of-samples of different sizes as scholars do when the true poverty is not observed.

\section{Comparing poverty prediction models}

\subsection{Baseline comparisons}

We compare the performance of classic econometric and machine learning models using eight models: The welfare and poverty models already described, and random forest, elastic net and neural network models each with continuous and dichotomous (categorical) dependent variable. We label these models wcn (Welfare - Continuous), rcn (Random Forest - Continuous), ecn (Elastic Net - Continuous), ncn (Neural Network - Continuous), pct (Poverty - Categorical), rct (Random Forest - Categorical), ect (Elastic Net - Categorical) and nct (Neural Network - Categorical) where `w' stands for welfare, `p' for poverty, `r' for random forest, `e' for elastic net, `n' for neural network, `cn' for continuous and `ct' for categorical model. This allows us comparing the performance of econometric and machine learning models and also the performance of continuous and dichotomous dependent variable models. The full results of the baseline OLS and Logit (Step 1) models are provided in Tables \ref{table:Table 3 - ols} and \ref{table:Table 4 - logit}.

\begin{center}
[Tables \ref{table:Table 3 - ols} and \ref{table:Table 4 - logit}]
\end{center}

All models are estimated in Stata with the following commands: `regress' (OLS-wcn), `logit' (Logit-pct), `rforest' (Random Forest - rcn and rct), `elasticnet' (Elastic Net - ecn and ect), and `mlp2' (Neural Network - ncn and nct). We are predicting all incomes using the full observed distribution of incomes for the modeling (step 1) equation. For all models, we use the same poverty line set at median income and the same set of explanatory variables with no interactions between variables\footnote{Some models such as random forest will work in a way that amounts to interacting variables but this is not done by design with the inclusion of interactions variables among regressors.}, we do not use any kind of weight and we do not use clustering of standard errors or any other estimation options. Alternative specifications of the models as well as alternative choice of poverty lines and models' options are discussed further in the paper.

The baseline models used in this section provide the simplest of the specifications required by the Stata routines. For the OLS and Logit models specifications are reported in Tables \ref{table:Table 3 - ols} and \ref{table:Table 4 - logit}. The same set of regressors is used for all models, including machine learning models. The hyperparameters of the Random forest model are set at 100 iterations (trees) with a default depth of 0, and a mtry parameter (the number of input variables at each iteration) equal to the squared root of the number of independent variables. The main hyperparameters for the Elastic Net model are $\alpha =0$ with the number of grid points set for $\lambda$ equal to 100. The main hyperparameters for the Nerural Network model are: 1) the number of neurons in the 1st and 2nd hidden layer are set at 100; 2) no bias term; 3) optimizer is a generic stochastic gradient descent; 4) the loss function is softmax for categorical dependent variables and MSE for continuous variables; 5) the initializing variance factor is 1; 6) the max number of restarts is 10; 7) the learning rate of the optimizer is 0.1; 8) the gradient smoothing term is 1e-8; 9) the first hidden layer dropout probability is 0; 10) the training batch size is 50 or the entire sample; 11) the maximum number of iterations is 100; 12) the report loss values at every number of iterations is 0. The ML models are not optimized. The spirit of  Tables \ref{table:Table 5 - comptable1} and \ref{table:Table 6 - ResShareType} is to present the results of the default routines offered by the Stata packages to mimic what most non-machine learning poverty specialist would do when they use ML models. This can be described as a baseline approach, which is a rather common approach among practitioners due to its computational speed and simplicity. The fine tuning of machine learning models with a comprehensive grid search is provided under the model's calibration section below.

Table \ref{table:Table 5 - comptable1} compares these baseline models. The top of the table reports the number of observations, the true poverty rate set at 50\%, predicted poverty rates, the difference and the t-tests for means difference between the true and predicted poverty rates. We then report two alternative functions where we give larger preference for TP and TN respectively (prefTN with $a=1.25$ and $b=0.75$; prefTP with $a=0.25$ and $b=1.25$). The rest of the indicators are the values for the four cells of the confusion matrix and the objective functions popular across the social sciences including Leakage, Undercoverage, Sensitivity, Specificity, Precision and Accuracy. As discussed, each indicator gives different weights to the different cells of the confusion matrix. In parenthesis (min/max), it is indicated whether the objective function is to be minimized or maximized. The table also reports the ranking of models based on each objective function (`1' is the best model and `8' is the worse).

\begin{center}
[Table \ref{table:Table 5 - comptable1}]
\end{center}

The table shows that random forest with categorical variable (rct) outperforms all other models according to all objective functions with the exception of the difference between true and predicted poverty rate where random forest with continuous dependent variable (rcn) comes out on top. Irrespective of the objective function, random forest outperforms all other models. It is also evident that a simple OLS model with no error term correction performs rather poorly because of its inability to predict incomes on the tails well.

To better understand what determines these findings, it is useful to plot the distributions of predicted values in the continuous case and the distributions of predicted probabilities in the dichotomous case. Figure 2 shows that the different models perform relatively well in some parts of the distributions but not in others. For the continuous case, all models tend to better perform around the center of the distribution and much less on the tails with the OLS and Elastic Net models being particularly poor on the tails and the Neural Network model being better on the lower tail but very poor on the upper tail. The only model that performs well around the middle of the distribution and has a relatively better performance on the tails is the random forest model. For the dichotomous case, we see similar patterns for all models except the random forest model. This model seems to be better able to split the poor and non-poor into two clearly distinguishable groups as opposed to the other models which have a large area of predictions that could easily switch between poor and non-poor depending on the probability cut-point chosen (an arbitrary choice of 50\% in our case. Other thresholds including optimized thresholds are considered later in the paper). Again, the random forest model would seem less susceptible to changes in the distribution of incomes and also to the probability cut point chosen to split the poor and the non-poor. All this keeping in mind that none of the machine learning model is fine tuned through grid search, something that is done later in the paper.

\begin{center}
[Figure 2]
\end{center}

\subsection{Missing data shares and patterns with alternative poverty lines}

Table \ref{table:Table 6 - ResShareType} compares predicted poverty rates across the different models using different missing data shares and patterns as described in the data section and different poverty lines. Note that, by imposing different shares and patterns of missing observations, we are also dictating the partition of observations between in-sample (observed incomes used for training the model) and out-of-sample (unobserved incomes used for testing the model). We are also testing, therefore, how models behave when the partition between train and test samples changes. This is particularly relevant for machine learning models.\footnote{There are more sophisticated methods to partition the train and test samples such as ``upsampling''. Given the variety of shares and patterns of missing observations tested in this section, we will not discuss or use alternative methods.} We test these models against four poverty lines: 5\%, 25\%, 50\%, and 75\% of the income distribution. This captures most of the poverty rates found across the world using the World Bank 2.15, 3.65, and 6.85 USD 2017 Purchasing Power Parity (PPP) poverty lines.\footnote{See https://pip.worldbank.org/home}.

\begin{center}
[Table \ref{table:Table 6 - ResShareType}]
\end{center}

Random forest with the dichotomous dependent variable (rct) emerges on top for most but not all estimations. For a poverty line of 25\% random forest is unambiguously the best model, on average across missing data patterns, and performs very well for most of the other poverty lines. It is also the top performing model for the most common and problematic of the scenarios (MAR-MNAR) with the exception of when the poverty line is set at 75\%.  It is also the case that random forest with continuous variable (rcn) comes into second place across models overall. However, there is no consistent dominance of random forest models either across all poverty lines or across all missing data patterns. For example, with a poverty line of 75\%, random forest comes out on top in only half of the missing data patterns.\footnote{It is important to stress here that ML models are not adapted to the size of the train sample. Some models like neural network require a minimum size for the train sample set whereas other models need to adapt the choice of parameters to the train sample size. These aspects are ignored here and we should consider that ML models can be improved as shown further in the paper. For the welfare and poverty models, we also tested multiple imputations methods as opposed to single imputation (results omitted). Results do not vary as we should expect. Values predicted with multiple imputation are means across repeated samples with replacement and these means center around the sample mean obtained with single imputation. What multiple imputation does is to improve on the estimate of the standard error, which can be larger or smaller than the standard error obtained with single imputation depending on the specification of the model. However, it does not improve on the predicted central value.} As compared to Table \ref{table:Table 5 - comptable1}, Table \ref{table:Table 6 - ResShareType} provides the important additional result that models' performance can depend on the poverty line and the missing data pattern.
 
Looking at the Cumulative Distribution Functions of predicted values helps us understanding the results (Figure 3. It is clear that CDFs cut across each other and some also cut across the original income distribution (left-hand panel). This means that there is no absolute dominance along the distribution with some models predicting income or poverty consistently lower or higher than other models or the original income distribution. Models that may perform better with low poverty lines (high probability thresholds) may not perform better with high poverty lines (low probability thresholds) and vice-versa. Predicted values are the further away from the original distribution on the tails explaining why poverty predictions are far from the true values when we use poverty lines that approach the 25th or 75th percentile. However, random forest is an exception in this respect as the CDF of predicted values is the closest of all models to the tails of the true income distribution in the continuous case. In the dichotomous case, random forest shows the most extreme curve on the tails indicating that its mass is concentrated closer to 0 and 1 probabilities. This makes classification of observations easier and more accurate with less observations located around the probability cutpoint.

\begin{center}
[Figure 3]
\end{center}

\section{Models' calibration}
So far, we have been using all models with the most basic specifications and tuning as provided by the Stata packages we used. In this section, we test how varying models' specifications and parameters can affect the estimation of poverty. 

\subsection{Specification of the modeling equation}
The set of independent variables and the explanatory power that this set determines may also be a discriminatory factor for the choice of an optimal prediction model. If an important variable is not included into the prediction equation, none of the models will benefit from this variable. However, machine learning models have the ability to improve on the use of the existing set of independent variables by including or omitting variables, or interacting them. Therefore, the initial set of independent variables may benefit some models more than others. 

In Table \ref{table:Table 7 - ExplanPower} we compare the initial set of independent variables we used thus far (Model1) with three other sets (Models2, Model3 and Model4). Model2 uses the same variables as model1 but changes the order by placing the continuous dependent variables at the bottom rather than at the top. Model 3 is a reduced model that includes all original variables as in Model1 except age and hhsize, two important predictors in our model. This reduces the R2 of the welfare OLS model from 0.33 to 0.22 and the Pseudo-R2 of the Logit poverty model from 0.18 to 0.11. Model4 is an even smaller model that keeps only the variables male, marital status and urban from Model1, which reduces the R2 of the OLS model to 0.12 and the Pseudo-R2 of the Logit model to 0.06. We can therefore compare four models with different orders of the independent variables and different sets of explanatory variables. The poverty line is set at 50\% by design as in Table \ref{table:Table 5 - comptable1}.

\begin{center}
[Table \ref{table:Table 7 - ExplanPower}]
\end{center}

A first obvious result is that, with the largest model (Model1), all models tend to perform better whereas none of the models performs well with the most reduced of the models (Model4). With the most parsimonious of the models (Model4), differences in poverty estimations across models almost disappear. With only three variables that explain little of the outcome variance, there is not much that machine learning models can add to simple OLS or logit models unless the model that best fits the data is not a linear model. 

An additional observation is that with an intermediary set of variables (Model3), the comparative advantage of random forest models disappears as compared to the more classical logit model, although it holds up well among the continuous dependent variable models. With parsimonious models, one may want to stick to a simple logit model. However, ML models can also be optimized for models with few predictors, which may improve further on the observed performance of the ML models we considered. 

Also noticeable is the fact that changing the order of variables changes results for the neural network model even if seeds are set for the random choices. This is possibly explained by the order in which neural network models choose the set of independent variables at each iteration. It is a known and highly debated issue among specialists. For the purpose of this paper and poverty analysis, it is important to keep in mind that changing the order of variables may affect the results of neural network models.

\subsection{OLS model (wcn)}
The real shortcoming of a linear regression model is its inability to predict incomes on the tails correctly. This is simply a statistical fact of OLS models which results in distributions of predicted values that are narrower than the original distributions, particularly if the explanatory power of the model is low. As surprising as it may seem, this is a problem that is routinely ignored in empirical works. One strand of the poverty literature that focused on this problem is the cross-survey imputation literature which proposed to address it by correcting the error term. In essence, the error term can be split into an idiosyncratic error term and a model error term. By estimating the model error term using the original empirical distribution or a theoretical normal distribution, one can add this error back into the predicted values mimicking in this way the variance of the original distribution. This is what we do in this section replicating the same technique used in cross-survey imputations. 

Results are shown in Table \ref{table:Table 8 - ComparingErrorAdjusted} providing corrections using the empirical distribution for the continuous dependent variable model and the logit model for the categorical dependent variable model. This form of imputation improves results substantially for all poverty lines bringing estimations much closer to the true values. It also outperforms other (non optimized) machine learning methods with the exception of random forest with a poverty line of 50\%. In the presence of missing data, particularly when the share of missing data is very large, and when the true poverty rate is expected to be far from the center of the distribution, it is essential to use this method when estimating poverty with OLS models. This simply confirms a fact that is very well known among cross-survey imputation specialists (\citealp{DangAl2019}). However and in our knowledge, one drawback of this method is that it is designed to estimate the poverty rate at the population level and does not provide imputations at the individual or household level. This means that one cannot use the objective functions derived from the confusion matrix. In other words, one can compare results with other models using only the predicted poverty rate. For this reason, this method could not be used to compare the full range of objective functions as we did in Tables \ref{table:Table 5 - comptable1}.

\begin{center}
[Table \ref{table:Table 8 - ComparingErrorAdjusted}]
\end{center}

\subsection{Probability threshold for dichotomous dependent variable models (pct, rct, ect, nct)}
It is possible to improve on predictions generated by dichotomous dependent variable models by optimizing the probability cutpoint used to separate the predicted poor from the predicted non-poor when households are classified after the prediction step. As probabilities of being poor vary between 0 and 1, most scholars use a cutpoint of 0.5 for simplicity, which is what we used so far. However, research across the social sciences has shown that one can use Receiver Operating Characteristics (ROC) curves and the Youden index (defined as y=max(sensitivity+specificity-1), or the max vertical distance between the ROC curve and the chance line) to optimize the cutpoint (see \citealp{Verme_2019} for a detailed discussion.). This should be of interest to poverty specialists who value sensitivity and specificity equally and wish to have a criteria to value the trade-off between these two objectives.

Table \ref{table:Table 9 - ProbThresholds} shows poverty rates for the four dichotomous dependent variable models, a range of reasonable cutpoints (between 45 and 55)\footnote{The Youden index results in cutpoints that are very rarely outside these boundaries}, and different poverty lines (5, 25, 50, and 75\%). We see that random forest outperforms all other models resulting in poverty predictions that are closer to the true poverty rate (5, 25, 50, and 75\%) with the exception of when the poverty rate is set at 50\% with a cutpoint of 0.45. If the objective function requires equal weight to sensitivity and specificity, maximizing the Youden index and finding the optimal cutpoint can improve poverty prediction for any model, but changing the cutpoint is not necessarily a discriminant for the choice of optimal model. 

\begin{center}
[Table \ref{table:Table 9 - ProbThresholds}]
\end{center}

\subsection{Fine tuning machine learning models (rcn, rct, ecn, ect, ncn, nct)}
Can machine learning models be improved for poverty predictions? The real strength of machine learning models is the set of parameters that can be tuned to optimize outcomes. This can be done with a trial and error approach or, more systematically, with a grid search approach. The latter consists of attributing a range of values to each parameter of the model, test all possible combinations of parameters, and compare results based on the objective function selected. This is the approach followed here. We conduct a grid search for each of the three machine learning models used in this paper and we compare results using the accuracy function ((TP+TN)/(TP+FP+TN+FN)*100), which is the most relevant function for a poverty analysis if the true poverty rate is not known. We split the data randomly into two equal samples and use sample 1 as training sample and sample 2 as test sample. The grid search for each model is conducted using the training sample and predictions are made using the test sample. Results for accuracy in the test sample are then compared across the three machine learning models. Table \ref{table:Table 10 - MachineLearninModels} provides the parameters' range and optimal values found through grid search.

\begin{center}
[Table \ref{table:Table 10 - MachineLearninModels}]
\end{center}

Results (max, mean, and standard deviation of accuracy) are shown in Table \ref{table:Table 11 - ComparingAccuracy} for three poverty lines (25, 50 and 75\%). It shows that grid search is effective in improving all models. Random forest remains the best model with a continuous dependent variable and ranks second with a categorical variable after the neural network model (see max values) . Overall and for each poverty line, all three models are able to reach rather high accuracy values (between 70 and 80\%) and also converge to a very narrow set of values across models. This means that all machine learning models are capable of providing good predictions provided that the parameters are finely tuned. The limits of the prediction capacity of these models seem to relate to the original input variables rather than the models themselves. The drawback of this approach is that it is extremely time consuming. While random forest was already close to optimal with a simple default specification of parameters and a run time of a few seconds, elastic net and neural network models required a full grid search to reach optimal results. This exercise eventually run into tens of hours of computing time explained by the large number of parameters that can be tuned, the vast number of values that can be attributed to each parameter, and the trial and error approach that one needs to follow to find a sensible range of values for each parameter. Based on these results, random forest remains the model of choice for poverty predictions, particularly with continuous dependent variable models.

\begin{center}
[Table \ref{table:Table 11 - ComparingAccuracy}]
\end{center}

\section{Concluding Remarks}
The paper has provided a comparative analysis of classic econometric and machine learning models used for the estimation of poverty at the population or household level in the presence of missing data. We carried out an artificial experiment using a dummy data set constructed on real data comparing eight different models and testing the robustness of results to changes in data, parameters and preferences. This strategy allowed us to address two important shortcomings of poverty prediction models: Test how different models perform with different missing data patterns including MNAR, and compare results with the true poverty rate. Below we provide some indications that can help practitioners consider alternative models for poverty predictions. These indications are preliminary, and some may be explained by a naive use of the models, but they can help to orient practitioners in the use of these models and scholars to structure future research. 

\begin{itemize}
\item Missing observations should always be of concern for poverty predictions unless they are a very low share of observations. Some models are not effective in predicting poverty even if missing observations are Missing Completely At Random (MCAR). 

\item Overall and for poverty predictions, no model can be expected to outperform all other models under any circumstance. The paper showed that models' relative performance can depend on the original distribution of incomes, the poverty line, models' parameters, the pattern of missing observations, the objective function and policy preferences. No model ``dominates'' others and predict incomes that are closer to the original distribution of incomes all along the distribution and in all cases tested.

\item The random forest model has proved to be the most consistent in predicting poverty relatively well under almost any condition considered in this paper. This is also consistent with tests conducted on other indicators of deprivation (\citealp{Andree2020}). It makes this model the most flexible and a preferred candidate when researchers lack key information for making a choice among models such as information about missing data patterns. For these models, it is important to have a sufficiently large number of iterations (trees) to have stable predictions and a proper depth, which may vary from case to case and needs to be tested with out-of-sample predictions.

\item Simple OLS models are generally ineffective in predicting poverty accurately if the model error term of predicted values is not adjusted post-estimation and the true poverty rate is distant from the mean. That is because of the narrow distribution of predicted values as compared to the original distribution (a statistical fact) and the incapacity of these models to predict income on the tails well.

\item OLS models can be substantially improved if the model error term is adjusted post-estimation as it is done in the cross-survey imputation literature. The drawback of this approach is that it provides estimates for only the poverty rate estimated at the population level rather than the individual or household level. As a consequence, it cannot be used to estimate all the objective functions derived from the confusion matrix.

\item Dichotomous dependent variable models tend to perform better, on average, than continuous models. This may be due to the probabilistic nature of predictions and their ability to fit the tails of an income distribution. However, this superiority should not be given for granted and does not apply when comparisons are made with OLS error adjusted models. Dichotomous dependent variable models can be improved by searching for the optimal probability cutpoint using ROC curves, but only marginally.

\item Baseline elastic net models with no fine tuning underperform relatively to logit and random forest models. These models are known to perform well with high dimensional regressors and, with more regressors, we should expect them to improve relatively to other models. A proper grid search analysis can significantly improve estimations coming very close but not always surpassing random forest models. A proper and comprehensive grid search comes with a significant cost in terms of computer time.  

\item Neural network models are very complex, time consuming and not easy to stabilize. We found that the baseline version of this model significantly underperforms other models. However, with a proper and comprehensive grid search one can reach optimal results that come close to random forest in the continuous dependent variable case, and surpass random forest in the categorical dependent variable case. As for the elastic net models, this result is achieved at a considerable cost in terms of computer time. Moreover, changing the order of variables may change results, an issue little known to poverty economists and not fully understood my ML specialists.

\item In the case of extremely reduced models with few independent variables (say 2-4 variables), there is not much difference in what model is used. All models will perform equally poorly. With a large set of independent variables (say more than 15), machine learning models have a comparative advantage in that they can test alternative reduced models and find the most effective in predicting poverty sparing researchers a complex and time consuming trial and error process.

\item The question of why random forest models perform relatively better than other models under most circumstances needs to be further studied. RF models are known to adapt well to non-linear relations which is one of the shortcomings of OLS models. But explaining the mechanics of this finding is not simple. One hypothesis is that the two random processes at work, random choice of sample and random choice of regressors, when combined exhaust the realm of possible predictions which results in better predictions on the tails. 

\end{itemize}

To conclude, in the absence of complete information on data, parameters and preferences, and in the absence of a deep understanding of machine learning models, simple logit models and random forest models should be preferred to simple OLS and other machine learning models. With time availability and a more in-depth knowledge of machine learning models, it would be important to try alternative machine learning models. The use of simple OLS models is not recommended whereas the OLS error adjustment method proposed by the cross-survey literature can be a very effective method to make OLS poverty predictions robust at the population level.

\bibliographystyle{chicago}
\bibliography{PredictingPoverty1}

\newpage

\begin{figure}[htbp]
	\centering
	\caption{Distributions of Income with Missing Data Patterns}
		\includegraphics[width=5.2in]{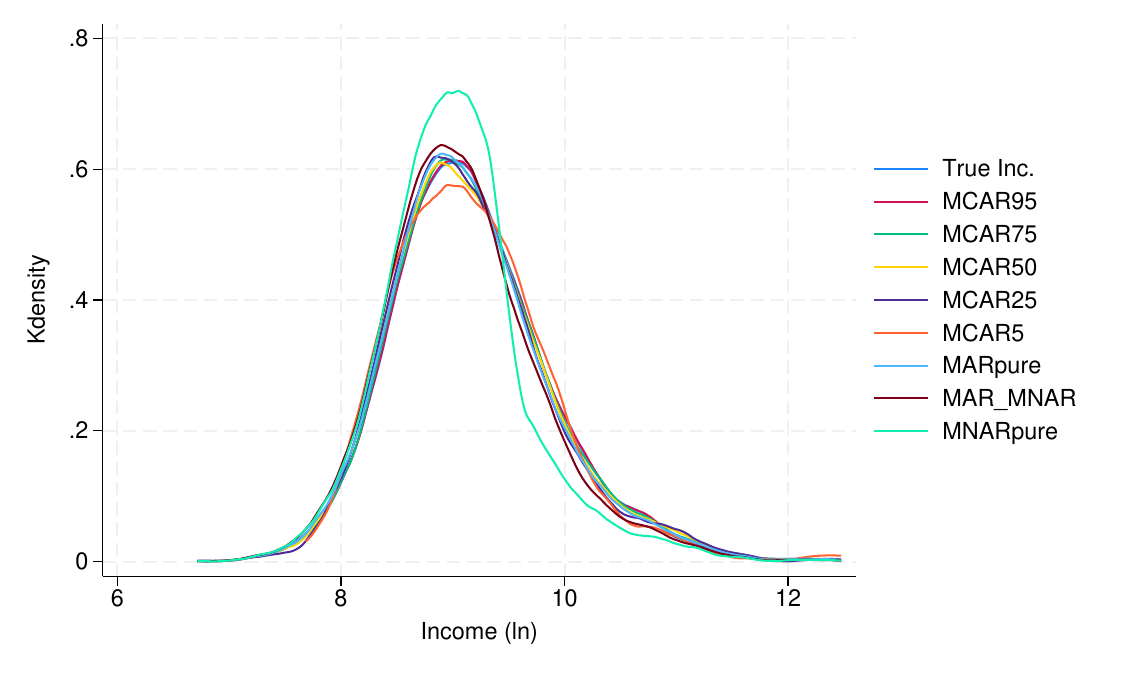}
	\label{fig:missdist}
	
{\raggedright \footnotesize \textbf{Legenda:} MCAR95-MCAR5 indicate the share of observed incomes with Missing Completely at Random observations. For example, with MCAR95, 95\% of incomes are observed and 5\% are missing randomly. MAR pure indicate observations Missing at Random only, MAR-MNAR indicate observations Missing at Random and Missing Not at Random; MNAR pure indicate observation Missing Not at Random only. See text for more details. \par}
\end{figure}

\newpage

\begin{figure}[hbt!]
	\centering
	\caption{Distributions of Predicted Values}
		\includegraphics[width=4.9in]{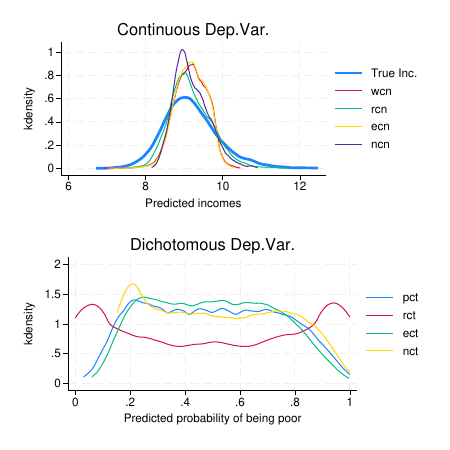}
	\label{fig:contcat}

{\raggedright \footnotesize \textbf{Legenda:} wcn (Welfare - Continuous); rcn (Random Forest - Continuous); ecn (Elastic Net - Continuous); ncn (Neural Network - Continuous); pct (Poverty - Categorical); rct (Random Forest - Categorical); ect (Elastic Net - Categorical) and nct (Neural Network - Categorical).  \par}
\end{figure}

\newpage

\begin{figure}[hbt!]
	\centering
	\caption{Cumulative Distributions of Predicted Values (CDFs)}
		\includegraphics[width=4.9in]{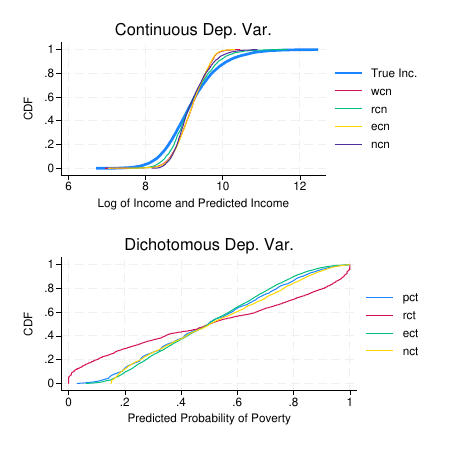}
	\label{fig:cumul}
\end{figure}

\newpage

\begin{table}[ht!]
\centering
\caption{True and Predicted Poverty Confusion Matrix}
\includegraphics[width=150mm]{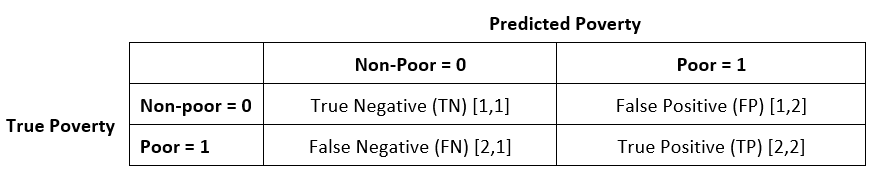}
Note: [x,y] indicates row and column.
\label{table:Table 1 - ConfusionMatrix}
\end{table}

\newpage

\begin{table}[h]
\centering
\caption{Summary Statistics}
\vspace*{5mm}
\begin{tabular}{lccccc}
\hline
\textbf{Variable}  & \textbf{Obs} & \textbf{Mean} & \textbf{Std. Dev.} & \textbf{Min} & \textbf{Max} \\
\hline 
household income             & 7,062        & 56887         & 54561              & 2800         & 1113157      \\
income per capita                & 7,062        & 13118         & 15098              & 826          & 261621       \\
age                & 7,062        & 51.64         & 14.00              & 15           & 98           \\
age squared               & 7,062        & 2863          & 1526               & 225          & 9604         \\
household size             & 7,062        & 5.14          & 2.43               & 1            & 24           \\
male               & 7,062        & 0.82          & 0.38               & 0            & 1            \\
marital status            & 7,062        & 0.83          & 0.38               & 0            & 1            \\
skills             & 7,062        & 0.19          & 0.39               & 0            & 1            \\
urban              & 7,062        & 0.60          & 0.49               & 0            & 1            \\
work\_salaried     & 7,062        & 0.39          & 0.49               & 0            & 1            \\
work\_selfemployed & 7,062        & 0.31          & 0.46               & 0            & 1            \\
work\_unpaid       & 7,062        & 0.00          & 0.05               & 0            & 1            \\
econ.sect.\_secondary          & 7,062        & 0.17          & 0.37               & 0            & 1            \\
econ.sect\_tertiary         & 7,062        & 0.33          & 0.47               & 0            & 1            \\
out of labor force        & 7,062        & 0.26          & 0.44               & 0            & 1     
\\
\hline
\end{tabular}

\label{table:Table 2 - SummaryStatistics}
\end{table}
Source: Author's estimations.

\newpage

\begin{table}[h]
\centering
\caption{OLS Baseline Regression (Dep.Var=HH Inc./Cap.)}
\vspace*{5mm}
\begin{tabular}{l*{4}{c}}
\hline
            &           b&          se&           t&      pvalue\\
\hline
age         &    .0165937&     .003265&    5.082352&    3.82e-07\\
age2        &   -.0001052&    .0000301&   -3.495455&    .0004762\\
hhsize      &   -.1087016&    .0031699&   -34.29193&    1.7e-238\\
male        &    .1536354&    .0297315&    5.167423&    2.44e-07\\
marstat     &   -.0538351&    .0289874&   -1.857187&    .0633262\\
skills      &    .5101445&    .0207039&    24.64005&    1.2e-128\\
urban       &    .2795772&    .0180025&    15.52991&    1.66e-53\\
work\_salaried&   -.4529071&    .0386937&   -11.70494&    2.34e-31\\
work\_selfemployed&   -.3464273&    .0395114&   -8.767785&    2.26e-18\\
sect\_sec    &   -.0749966&    .0264836&   -2.831814&    .0046416\\
sect\_tert   &    .0828654&    .0233992&    3.541374&    .0004006\\
out\_labor   &   -.1971939&    .0439236&   -4.489479&    7.25e-06\\
\_cons      &    9.156962&    .0929127&    98.55447&           0\\
\hline
\end{tabular}

\label{table:Table 3 - ols}
\end{table} 
Source: Author's estimations.

\newpage

\begin{table}[h]
\centering
\caption{Logit Baseline Regression (Dep.Var.=HH Poor/Non-Poor)}
\vspace*{5mm}
\begin{tabular}{l*{4}{c}}
\hline
            &           b&          se&           z&      pvalue\\
\hline
truepoor    &            &            &            &            \\
age         &      -.0465&    .0128019&   -3.632287&    .0002809\\
age2        &    .0003321&    .0001182&    2.809007&    .0049695\\
hhsize      &    .3561425&    .0147467&    24.15058&    7.4e-129\\
male        &   -.3922606&     .114842&   -3.415653&    .0006363\\
marstat     &    .0242854&     .112487&    .2158953&    .8290693\\
skills      &   -1.319658&    .0817895&   -16.13481&    1.45e-58\\
urban       &   -.7173156&    .0685802&   -10.45952&    1.33e-25\\
work\_salaried&    1.297566&    .1604475&    8.087167&    6.11e-16\\
work\_selfemployed&    .9322508&    .1616109&    5.768491&    8.00e-09\\
sect\_sec    &    .0441941&    .1030088&    .4290316&    .6679002\\
sect\_tert   &   -.3580383&    .0898093&   -3.986649&     .000067\\
out\_labor   &    .3610466&    .1765711&    2.044765&     .040878\\
\_cons      &   -.1580234&    .3644101&   -.4336416&    .6645487\\
\hline
\end{tabular}

\label{table:Table 4 - logit}
\end{table} 
Source: Author's estimations.

\newpage

\begin{table}[htb!]
\centering
\caption{Comparing Poverty Prediction Models Using Different Objective Functions }
\vspace*{5mm}
\resizebox{1\textwidth}{!}{
\begin{tabular}{l*{16}{c}}
\hline
            &         wcn&       wcn\_r&         rcn&       rcn\_r&         ecn&       ecn\_r&         ncn&       ncn\_r&         pct&       pct\_r&         rct&       rct\_r&         ect&       ect\_r&         nct&       nct\_r\\
\hline
Observations&        7062&           .&        7062&           .&        7062&           .&        7062&           .&        7062&           .&        7062&           .&        7062&           .&        7062&           .\\
TruePovRate &          50&           .&          50&           .&          50&           .&          50&           .&          50&           .&          50&           .&          50&           .&          50&           .\\
PredPoverty &       43.09&           .&       50.06&           .&       43.16&           .&       49.63&           .&       49.11&           .&       49.92&           .&       49.45&           .&       50.48&           .\\
Diff.(absmin)&        6.91&           8&         .06&           1&        6.84&           7&         .37&           3&         .89&           6&         .08&           2&         .55&           5&         .48&           4\\
Diff.(tstat)&       10.61&           .&        -.12&           .&       10.51&           .&          .6&           .&        1.39&           .&          .2&           .&         .86&           .&        -.74&           .\\
PrefTruePos(max)&       67.83&           8&       83.08&           2&       67.94&           7&       72.95&           3&       70.76&           4&       87.66&           1&       70.68&           5&       70.58&           6\\
PrefTrueNeg(max)&       71.28&           5&       83.05&           2&       71.36&           4&       73.13&           3&       71.21&           6&        87.7&           1&       70.95&           7&       70.34&           8\\
TruePos(max)&        2212&           8&        2935&           2&        2218&           7&        2566&           3&        2475&           6&        3093&           1&        2481&           5&        2505&           4\\
TrueNeg(max)&        2700&           4&        2931&           2&        2701&           3&        2592&           5&        2538&           6&        3099&           1&        2520&           7&        2471&           8\\
FalsePos(min)&         831&           4&         600&           2&         830&           3&         939&           5&         993&           6&         432&           1&        1011&           7&        1060&           8\\
FalseNeg(min)&        1319&           8&         596&           2&        1313&           7&         965&           3&        1056&           6&         438&           1&        1050&           5&        1026&           4\\
Leakage(min)&       23.53&           4&       16.99&           2&       23.51&           3&       26.59&           5&       28.12&           6&       12.23&           1&       28.63&           7&       30.02&           8\\
Undercoverage(min)&       37.35&           8&       16.88&           2&       37.18&           7&       27.33&           3&       29.91&           6&        12.4&           1&       29.74&           5&       29.06&           4\\
Sensitivity(max)&       62.65&           8&       83.12&           2&       62.82&           7&       72.67&           3&       70.09&           6&        87.6&           1&       70.26&           5&       70.94&           4\\
Specificity(max)&       76.47&           4&       83.01&           2&       76.49&           3&       73.41&           5&       71.88&           6&       87.77&           1&       71.37&           7&       69.98&           8\\
Precision(max)&       72.69&           5&       83.03&           2&       72.77&           4&       73.21&           3&       71.37&           6&       87.74&           1&       71.05&           7&       70.27&           8\\
Accuracy(max)&       69.56&           8&       83.06&           2&       69.65&           7&       73.04&           3&       70.99&           4&       87.68&           1&       70.82&           5&       70.46&           6\\
\hline
\end{tabular}

\label{table:Table 5 - comptable1} 
}
\end{table} 
Source: Author's estimations. Legenda: wcn (Welfare - Continuous); rcn (Random Forest - Continuous); ecn (Elastic Net - Continuous); ncn (Neural Network - Continuous); pct (Poverty - Categorical); rct (Random Forest - Categorical); ect (Elastic Net - Categorical) and nct (Neural Network - Categorical). wcn-r refers to the rank position of the wcn model (horizontal ranking) with `1' indicating the top performing model and `8' the worse performing model. Similarly for other models. `Diff' refers to the difference between the true and the predicted poverty rates. Leakage=FP/(FP+TN); Undercoverage=FN/(FN+TP); Sensitivity=TP/(FN+TP); Specificity=TN/(TN+FP); Precision=TP/(TP+FP); Accuracy=(TP+TN)/N. (min) (max) indicate whether the objective function is to be minimized or maximized. (absmin) indicates that the absolute value is to be minimized.

\newpage

\begin{table}[hbt!]
\centering
\caption{Comparing Poverty Prediction Models Using Different Types and Shares of Missing Observations and Poverty Lines}
\vspace*{5mm}
\resizebox{1\textwidth}{!}{
\begin{tabular}{l*{16}{c}}
\hline
            &         wcn&       wcn\_r&         rcn&       rcn\_r&         ecn&       ecn\_r&         ncn&       ncn\_r&         pct&       pct\_r&         rct&       rct\_r&         ect&       ect\_r&         nct&       nct\_r\\
\hline
PovLine=5\% &           .&           .&           .&           .&           .&           .&           .&           .&           .&           .&           .&           .&           .&           .&           .&           .\\
MCAR95      &         4.8&           1&         4.8&           2&         4.8&           3&         4.7&           7&         4.8&           4&         4.8&           5&         4.8&           6&         4.7&           8\\
MCAR75      &         4.1&           2&           4&           3&           4&           4&         3.7&           6&         3.8&           5&         4.2&           1&         3.7&           7&         3.7&           8\\
MCAR50      &         2.9&           2&         2.7&           4&         2.8&           3&         2.5&           6&         2.6&           5&         3.2&           1&         2.5&           7&         2.5&           8\\
MCAR25      &         1.9&           2&         1.6&           4&         1.8&           3&         1.2&           7&         1.5&           5&         2.7&           1&         1.3&           6&         1.2&           8\\
MCAR5       &           1&           1&          .2&           6&          .9&           3&           1&           2&          .6&           5&          .7&           4&          .2&           7&          .2&           8\\
MARpure     &         3.2&           2&         2.9&           5&         3.2&           3&         2.9&           6&           3&           4&         3.4&           1&         2.9&           7&         2.9&           8\\
MAR\_MNAR    &         3.9&           2&         3.9&           3&         3.9&           4&         3.9&           5&         3.9&           6&           4&           1&         3.9&           7&         3.9&           8\\
MNARpure    &         4.2&           2&         4.2&           3&         4.2&           4&         4.2&           5&         4.2&           6&         4.3&           1&         4.2&           7&         4.2&           8\\
Average     &         3.3&         1.8&           3&         3.8&         3.2&         3.4&           3&         5.5&           3&           5&         3.4&         1.9&         2.9&         6.8&         2.9&           8\\
PovLine=25\%&           .&           .&           .&           .&           .&           .&           .&           .&           .&           .&           .&           .&           .&           .&           .&           .\\
MCAR95      &        24.5&           3&        24.6&           2&        24.4&           5&        23.7&           8&        24.4&           6&        24.8&           1&        24.4&           7&        24.5&           4\\
MCAR75      &        21.9&           3&        22.3&           2&        21.7&           4&        18.7&           7&        21.7&           5&        23.4&           1&        21.4&           6&        18.7&           8\\
MCAR50      &        18.2&           3&        20.1&           2&        17.6&           5&        13.5&           8&        17.6&           6&        21.9&           1&        16.7&           7&        18.1&           4\\
MCAR25      &        15.2&           3&        16.7&           2&        14.2&           5&         6.2&           8&        14.6&           4&        20.5&           1&        12.9&           6&        12.6&           7\\
MCAR5       &        10.5&           3&         5.6&           7&         9.4&           4&           8&           6&          11&           2&        13.6&           1&         4.3&           8&         8.3&           5\\
MARpure     &        18.6&           4&        20.2&           2&        18.4&           5&        15.9&           8&        18.7&           3&        21.8&           1&          18&           7&        18.1&           6\\
MAR\_MNAR    &        19.7&           4&        20.1&           2&        19.7&           5&        19.9&           3&        19.7&           6&        20.8&           1&        19.7&           7&        19.7&           8\\
MNARpure    &        21.4&           3&        21.8&           2&        21.4&           4&        21.3&           8&        21.4&           5&        22.2&           1&        21.4&           6&        21.4&           7\\
Average     &        18.8&         3.3&        18.9&         2.6&        18.4&         4.6&        15.9&           7&        18.6&         4.6&        21.1&           1&        17.4&         6.8&        17.7&         6.1\\
PovLine=50\%&           .&           .&           .&           .&           .&           .&           .&           .&           .&           .&           .&           .&           .&           .&           .&           .\\
MCAR95      &        49.6&           6&        49.9&           1&        49.6&           7&        47.5&           8&        49.9&           2&        50.1&           3&        49.9&           4&        49.9&           5\\
MCAR75      &        47.8&           6&        49.4&           4&        47.8&           7&        37.5&           8&        49.6&           1&        49.6&           2&        49.5&           3&        49.2&           5\\
MCAR50      &        45.7&           6&        48.8&           4&        45.5&           7&        29.5&           8&        49.2&           3&        49.7&           1&        49.7&           2&        47.3&           5\\
MCAR25      &        43.6&           6&        47.8&           5&        43.6&           7&        12.5&           8&        49.3&           2&          50&           1&        49.1&           3&        50.9&           4\\
MCAR5       &        44.3&           6&        49.2&           2&        44.9&           5&        35.1&           8&        52.1&           4&        50.2&           1&        51.8&           3&          44&           7\\
MARpure     &        44.4&           7&        47.8&           2&        44.3&           8&        52.3&           4&        47.2&           5&        49.3&           1&          47&           6&        47.8&           3\\
MAR\_MNAR    &        42.3&           6&        43.3&           2&        42.2&           7&        41.7&           8&        43.1&           4&        44.5&           1&        43.2&           3&        43.1&           5\\
MNARpure    &        44.4&           6&        45.5&           2&        44.4&           7&        42.6&           8&        44.5&           4&        46.2&           1&        44.5&           5&        44.8&           3\\
Average     &        45.3&         6.1&        47.7&         2.8&        45.3&         6.9&        37.3&         7.5&        48.1&         3.1&        48.7&         1.4&        48.1&         3.6&        47.1&         4.6\\
PovLine=75\%&           .&           .&           .&           .&           .&           .&           .&           .&           .&           .&           .&           .&           .&           .&           .&           .\\
MCAR95      &        75.3&           4&        75.1&           3&        75.4&           5&        76.2&           8&        75.4&           6&          75&           1&        75.5&           7&          75&           2\\
MCAR75      &        76.6&           3&        76.3&           2&        76.7&           4&        81.2&           8&        77.1&           5&        75.9&           1&        77.6&           7&        77.1&           6\\
MCAR50      &        78.4&           3&        77.5&           2&        78.7&           4&        87.5&           8&        79.5&           5&        76.7&           1&        80.7&           7&        79.7&           6\\
MCAR25      &        80.8&           3&        79.5&           2&        81.3&           5&        93.7&           8&        82.7&           6&        78.7&           1&        84.7&           7&          81&           4\\
MCAR5       &        89.7&           4&        87.9&           3&        90.8&           5&        91.7&           7&        87.4&           2&          84&           1&        91.1&           6&        98.7&           8\\
MARpure     &        76.6&           4&        75.7&           3&        76.7&           5&        78.1&           7&        77.6&           6&        75.4&           2&        78.4&           8&        75.2&           1\\
MAR\_MNAR    &        71.8&           3&        71.3&           7&          72&           2&        69.2&           8&        71.4&           4&        71.4&           5&        72.5&           1&        71.4&           6\\
MNARpure    &        71.1&           8&        71.3&           5&        71.3&           6&        78.7&           7&        72.8&           3&        71.8&           4&        73.7&           1&        73.5&           2\\
Average     &        77.5&           4&        76.8&         3.4&        77.9&         4.5&          82&         7.6&          78&         4.6&        76.1&           2&        79.3&         5.5&        78.9&         4.4\\
\hline
\end{tabular}

\label{table:Table 6 - ResShareType}
}
\end{table}
Source: Author's estimations. Legenda: wcn (Welfare - Continuous); rcn (Random Forest - Continuous); ecn (Elastic Net - Continuous); ncn (Neural Network - Continuous); pct (Poverty - Categorical); rct (Random Forest - Categorical); ect (Elastic Net - Categorical) and nct (Neural Network - Categorical).  wcn-r refers to the rank position of the wcn model (horizontal ranking) with `1' indicating the top performing model and `8' the worse performing model. Similarly for other models. MCAR95-MCAR5 indicate the share of observed incomes with Missing Completely at Random observations. MAR pure indicate observations Missing at Random only, MAR-MNAR indicate observations Missing at Random and Missing Not at Random; MNAR pure indicate observation Missing Not at Random only. See text for more details.

\newpage

\begin{table}[hbt!]
\centering
\caption{Comparing Poverty Prediction Models Using Different Specifications}
\vspace*{5mm}
\resizebox{1\textwidth}{!}{
\begin{tabular}{l*{16}{c}} \hline
            &         wcn&       wcn\_r&         rcn&       rcn\_r&         ecn&       ecn\_r&         ncn&       ncn\_r&         pct&       pct\_r&         rct&       rct\_r&         ect&       ect\_r&         nct&       nct\_r\\
\hline
Model1      &        43.2&           7&        50.1&           1&        43.2&           8&          48&           6&        49.2&           5&        50.1&           2&        49.4&           3&        50.6&           4\\
Model2      &        43.2&           6&        50.1&           1&        43.2&           7&          43&           8&        49.2&           4&        50.3&           2&        49.4&           3&        54.5&           5\\
Model3      &        42.3&           8&        51.9&           3&          43&           7&        45.4&           6&        50.7&           1&        53.5&           5&        51.1&           2&        52.5&           4\\
Mobel4      &        39.6&           2&        39.6&           3&        39.6&           4&        38.4&           8&        39.6&           5&        39.6&           6&        39.6&           7&        41.8&           1\\
\hline
\end{tabular}

\label{table:Table 7 - ExplanPower}
}
\end{table}
Source: Author's estimations. Legenda: wcn (Welfare - Continuous); rcn (Random Forest - Continuous); ecn (Elastic Net - Continuous); ncn (Neural Network - Continuous); pct (Poverty - Categorical); rct (Random Forest - Categorical); ect (Elastic Net - Categorical) and nct (Neural Network - Categorical).  wcn-r refers to the rank position of the wcn model (horizontal ranking) with `1' indicating the top performing model and `8' the worse performing model. The poverty line is set at the Median value (50\%) by design.

\newpage

\begin{table}[hbt!]
\centering
\caption{Comparing Error Adjusted OLS model with Other Models for Different Poverty Lines}
\vspace*{5mm}
\begin{tabular}{lccc}
\hline
\multicolumn{1}{r}{Poverty Line (\%)} & 25                   & 50                   & 75                   \\
\textbf{Continuous Dep. Var.}         & \multicolumn{1}{l}{} & \multicolumn{1}{l}{} & \multicolumn{1}{l}{} \\
\hline
OLS                                   & 12.1                 & 43.2                 & 82.1                 \\
OLS povimp (empirical)                & 25.0                 & 47.4                 & 73.2                 \\
Random Forest                         & 16.6                 & 49.7                 & 81.0                 \\
Elastic Net                           & 11.1                 & 42.9                 & 83.0                 \\
Neural Network                        & 13.0                 & 47.5                 & 77.4                 \\
\textbf{Categorical Dep. Var.}        &                      &                      &                      \\
Logit                                 & 11.4                 & 51.9                 & 84.8                 \\
Logit povimp (empirical)              & 25.5                 & 51.0                 & 75.7                 \\
Random Forest                         & 20.2                 & 50.9                 & 79.7                 \\
Elastic Net                           & 10.1                 & 52.2                 & 87.3                 \\
Neural Network                        & 16.2                 & 54.9                 & 87.1                
\\
\hline
\end{tabular}

\label{table:Table 8 - ComparingErrorAdjusted}
\end{table}
Source: Author's estimations.

\newpage

\begin{table}[hbt!]
\centering
\caption{Comparing Dichotomous Poverty Prediction Models Using Different Probability Cutpoints and Different Poverty Lines}
\vspace*{5mm}
\begin{tabular}{l*{8}{c}} \hline
            &         pct&       pct\_r&         rct&       rct\_r&         ect&       ect\_r&         nct&       nct\_r\\
\hline
PL=5        &           .&           .&           .&           .&           .&           .&           .&           .\\
Cut=0\_45    &          .4&           2&        3.84&           1&         .23&           3&           0&           4\\
Cut=0\_50    &          .2&           2&        3.27&           1&         .16&           3&           0&           4\\
Cut=0\_55    &          .2&           2&        3.07&           1&         .11&           3&           0&           4\\
PL=25       &           .&           .&           .&           .&           .&           .&           .&           .\\
Cut=0\_45    &        15.4&           3&       25.15&           1&       13.89&           4&       18.58&           2\\
Cut=0\_50    &        11.5&           3&       21.75&           1&        9.52&           4&       14.27&           2\\
Cut=0\_55    &         7.8&           3&       18.82&           1&        6.22&           4&       10.96&           2\\
PL=50       &           .&           .&           .&           .&           .&           .&           .&           .\\
Cut=0\_45    &        55.4&           3&        54.6&           2&       56.24&           4&       53.65&           1\\
Cut=0\_50    &        49.2&           4&       50.14&           1&       49.39&           2&       50.64&           3\\
Cut=0\_55    &        43.9&           2&       46.05&           1&       42.69&           4&       42.78&           3\\
PL=75       &           .&           .&           .&           .&           .&           .&           .&           .\\
Cut=0\_45    &        87.1&           3&       79.35&           1&        89.8&           4&       85.58&           2\\
Cut=0\_50    &        84.2&           3&       77.37&           1&       86.12&           4&       82.82&           2\\
Cut=0\_55    &        78.3&           2&       74.64&           1&       81.05&           4&        79.3&           3\\
\hline
\end{tabular}

\label{table:Table 9 - ProbThresholds}
\end{table}

Source: Author's estimations. Legenda: pct (Poverty - Categorical); rct (Random Forest - Categorical); ect (Elastic Net - Categorical) and nct (Neural Network - Categorical).  pct-r refers to the rank position of the pct model (horizontal ranking) with `1' indicating the top performing model and `4' the worse performing model. Similarly for other models. PL=Poverty Line=True Poverty Rate. Cut=Probability cutpoint for separating predicted poor and non-poor.

\newpage

\begin{table}[hbt!]
\caption{Machine Learning Models: Grid Search Range and Optimal Parameters' Values}
\vspace*{5mm}
\centering
\begin{tabular}{lccccccc}
\hline
                                 & Grid range         & \multicolumn{6}{c}{Optimal parameters}                                               \\
                                 &                    & Cont.       & Cont.        & Cont.        & Cat.        & Cat.         & Cat.        \\
\textit{Poverty Line (\%)}       & \textit{}          & \textit{25} & \textit{50}  & \textit{75}  & \textit{25} & \textit{50}  & \textit{75} \\
\hline
\textit{\textbf{Random Forest}}  & \textit{\textbf{}} & \textit{}   & \textit{}    & \textit{}    & \textit{}   & \textit{}    & \textit{}   \\
Iterations                       & 50, 100, 200, 400  & 50          & 50           & 200          & 50          & 200          & 100         \\
Number of Vars                   & 1-12               & 6           & 9            & 9            & 6           & 3            & 11          \\
Depth                            & 3-8                & 8           & 6            & 5            & 7           & 7            & 6           \\
Leaf size                        & 5, 10, 50, 100     & 100         & 10           & 10           & 100         & 50           & 10          \\
\textit{\textbf{Elastic Net}}    & \textit{\textbf{}} &             &              &              &             &              &             \\
Alpha                            & 0, 2, 4, 6, 8, 1   & 0           & 0.2          & 0            & 1           & 0.2          & 0.8         \\
Lambda                           & 50, 100, 200       & 50          & 50           & 50           & 100         & 100          & 50          \\
Folds                            & 5, 10, 20          & 5           & 10           & 5            & 5           & 5            & 5           \\
\textit{\textbf{Neural Network}} & \textit{\textbf{}} & \textit{}   & \textit{}    & \textit{}    & \textit{}   & \textit{}    & \textit{}   \\
Layer 1                          & 64, 128, 256       & 128         & \textit{128} & \textit{256} & \textit{64} & \textit{128} & \textit{64} \\
Layer 2                          & 64, 128, 256       & 64          & 128          & 128          & 64          & 256          & 64          \\
Learning Rate                    & .01, .001          & 0.01        & 0.001        & 0.01         & 0.01        & 0.001        & 0.001       \\
Batch                            & 20, 80             & 20          & 20           & 80           & 20          & 20           & 20          \\
Epochs                           & 50, 200            & 50          & 200          & 50.0         & 50          & 50           & 50         
\\
\hline
\end{tabular}

\label{table:Table 10 - MachineLearninModels}
\end{table}
Source: Author's estimations.

\newpage

\begin{table}[hbt!]
\caption{Comparing Accuracy Scores in Test Samples Across Models}
\vspace*{5mm}
\centering
\begin{tabular}{lcccccc}
\hline
                           & Cont.       & Cont.       & Cont.       & Cat.        & Cat.        & Cat.        \\
\textit{Poverty Line (\%)} & \textit{25} & \textit{50} & \textit{75} & \textit{25} & \textit{50} & \textit{75} \\
\hline
\textbf{Max}               &             &             &             &             &             &             \\
\textit{Random forest}     & 78.5        & 71.4        & 80.7        & 78.3        & 70.9        & 80.2        \\
\textit{Elastic Net}       & 77.9        & 69.5        & 79.5        & 78.1        & 70.4        & 79.2        \\
\textit{Neural Network}    & 78.2        & 70.9        & 80.2        & 78.6        & 71.0        & 80.3        \\
\textbf{Mean}              &             &             &             &             &             &             \\
\textit{Random forest}     & 77.7        & 69.9        & 79.0        & 77.7        & 69.9        & 78.2        \\
\textit{Elastic Net}       & 77.8        & 69.4        & 79.4        & 78.0        & 70.4        & 79.2        \\
\textit{Neural Network}    & 77.2        & 70.0        & 79.3        & 78.1        & 70.2        & 79.2        \\
\textbf{Std.Dev.}          &             &             &             &             &             &             \\
\textit{Random forest}     & 0.6         & 0.9         & 1.4         & 0.4         & 0.8         & 1.5         \\
\textit{Elastic Net}       & 0.1         & 0.0         & 0.0         & 0.0         & 0.0         & 0.1         \\
\textit{Neural Network}    & 0.6         & 0.6         & 0.5         & 0.5         & 0.9         & 1.3   
\\
\hline     
\end{tabular}

\label{table:Table 11 - ComparingAccuracy}
\end{table}
Source: Author's estimations.

\end{document}